\documentclass[journal,twoside,web]{ieeecolor}
\usepackage{generic}
\usepackage{cite}
\usepackage{amsmath,amssymb,amsfonts}
\usepackage{algorithmic}
\usepackage{graphicx}
\usepackage{multirow}
\usepackage{comment}
\usepackage{multirow}
\usepackage{booktabs}
\usepackage{arydshln}
\usepackage{booktabs}

\usepackage{vcell}
\usepackage{textcomp}
\def\BibTeX{{\rm B\kern-.05em{\sc i\kern-.025em b}\kern-.08em
    T\kern-.1667em\lower.7ex\hbox{E}\kern-.125emX}}
\begin{document}
\title{HMRNet: High and Multi-Resolution Network with Bidirectional Feature Calibration for Brain Structure Segmentation in Radiotherapy}
    \author{Hao Fu, Guotai Wang, Wenhui Lei, Wei Xu, Qianfei~Zhao, Shichuan~Zhang, Kang~Li and Shaoting~Zhang
\thanks{
This work was supported by the National Natural Science Foundation of China under Grant 81771921 and Grant 61901084. (Corresponding authors: Guotai Wang; Shaoting Zhang.) }
\thanks{Hao Fu, Guotai Wang, Wenhui Lei, Wei Xu and Qianfei Zhao are with the School of Mechanical and Electrical Engineering, University of Electronic Science and Technology of China, Chengdu, 611731, China. Guotai Wang is also with Shanghai AI laboratory, Shanghai, 200030, China. (e-mail: guotai.wang@uestc.edu.cn)}
\thanks{Shichuan Zhang is with Department of Radiation Oncology, Sichuan Cancer Hospital and Institute, University of Electronic Science and Technology of China, Chengdu, 610042, China. 
}
\thanks{Kang Li is with West China Hospital-SenseTime Joint Lab, West China Biomedical Big Data Center, Sichuan University West China Hospital, Chengdu, 610041, China. 
}
\thanks{Shaoting Zhang is with the School of Mechanical and Electrical
Engineering, University of Electronic Science and Technology of China, Chengdu, 611731, China, and also with SenseTime Research, Shanghai, 200233, China.
(e-mail: zhangshaoting@uestc.edu.cn).
}}
\maketitle

\begin{abstract}
Accurate segmentation of Anatomical brain Barriers to Cancer spread (ABCs) plays an important role for automatic delineation of Clinical Target Volume (CTV) of brain tumors in radiotherapy.  Despite that variants of U-Net are state-of-the-art segmentation models, they have limited performance when dealing with ABCs structures with various shapes and sizes, especially  thin structures (e.g., the falx cerebri) that span only few slices. To deal with this problem, we propose a High and Multi-Resolution Network (HMRNet) that consists of a multi-scale feature learning branch and a high-resolution branch, which can maintain the high-resolution contextual information and extract more robust representations of  anatomical structures with various scales. We further design a Bidirectional Feature Calibration (BFC) block to enable the two branches to generate spatial attention maps for mutual feature calibration. Considering the different sizes and positions of ABCs structures, our network was applied after a rough localization of each structure to obtain fine segmentation results. Experiments on the MICCAI 2020 ABCs challenge  dataset showed that: 1) Our proposed two-stage segmentation strategy largely outperformed methods segmenting all the structures in just one stage; 2) The proposed HMRNet with two branches can maintain high-resolution representations and is effective to improve the performance on thin structures; 3) The proposed BFC block outperformed existing attention methods using monodirectional feature calibration. Our method won the second place of ABCs 2020 challenge and has a potential for more accurate and reasonable delineation of CTV of brain tumors. 
\end{abstract}

\section{Introduction}
\label{sec:introduction}
\IEEEPARstart{B}{rain}  cancer has became one of the most common cancers around the world \cite{bray2018global}. Radiotherapy plays an important role in clinical treatment of brain cancer, and is usually used before or after surgery \cite{bernier2004postoperative,bonner2010radiotherapy,jones2004treatment}. During radiotherapy treatment planning, accurate delineation of the Clinical Target Volume (CTV)~\cite{lai2011does} is very important for ensuring that the cancer cells receive sufficient radiation dose and that for healthy  structures lie in a safe range. However, the CTV is not radiographically distinguishable from healthy tissues, and in practice it is delineated by estimating the extent of microscopic disease spread based on accumulated knowledge of prior treatment outcomes and histological evidence on the extent of cancer cell spread~\cite{shusharina2020automated}. While the CTV itself is ambiguous, it is much easier to identify structures that do not belong to the CTV and serve as Anatomical Brain Barriers to Cancer spread (ABCs)~\cite{shusharina2020automated}. For gliomas,  the most common type of brain tumors, the CTV boundary is  constrained by barrier structures like the falx cerebri, tentorium cerebelli, brain sinuses, cerebellum and ventricles. Therefore, accurate segmentation of these barrier structures has a potential to assist more accurate and reasonable delineation of CTV that can improve the radiotherapy outcome of brain cancer patients.

Meanwhile, delineation of anatomical structures relies on multi-modal medical images, such as Computed Tomography (CT) and T1- and T2-weighted Magnetic Resonance  Imaging (MRI). CT is required by radiotherapy dose computation as it can measure the density of different structures. It can show dense structures like bones and implants with less distortion, but has a low contrast for soft tissues~\cite{bhatnagar2015new}. T1- and T2-weighted MRI can better highlight different soft tissues than CT~\cite{bhatnagar2015new}. Therefore, employing different imaging modalities can provide complementary information for accurate structure segmentation required by radiotherapy treatment planning. At present, the delineation of multiple brain structures is implemented by experienced radiation oncologists through slice-by-slice manual annotation, which is labor intensive, time consuming, and subject to the operator's experience. Thus, accurate automatic segmentation of the anatomical structures like ABCs is desirable for reproducibility and reducing the workload of annotators and the waiting time for patients~\cite{sharp2014vision}. Considering the state-of-the-art performance of Convolutional Nerual Networks (CNNs) in medical image analysis~\cite{litjens2017survey,shen2017deep} due to their powerful feature representations capability, it is promising to use CNNs to automate this task.  

\begin{figure}
\centering\includegraphics[width=1\columnwidth]{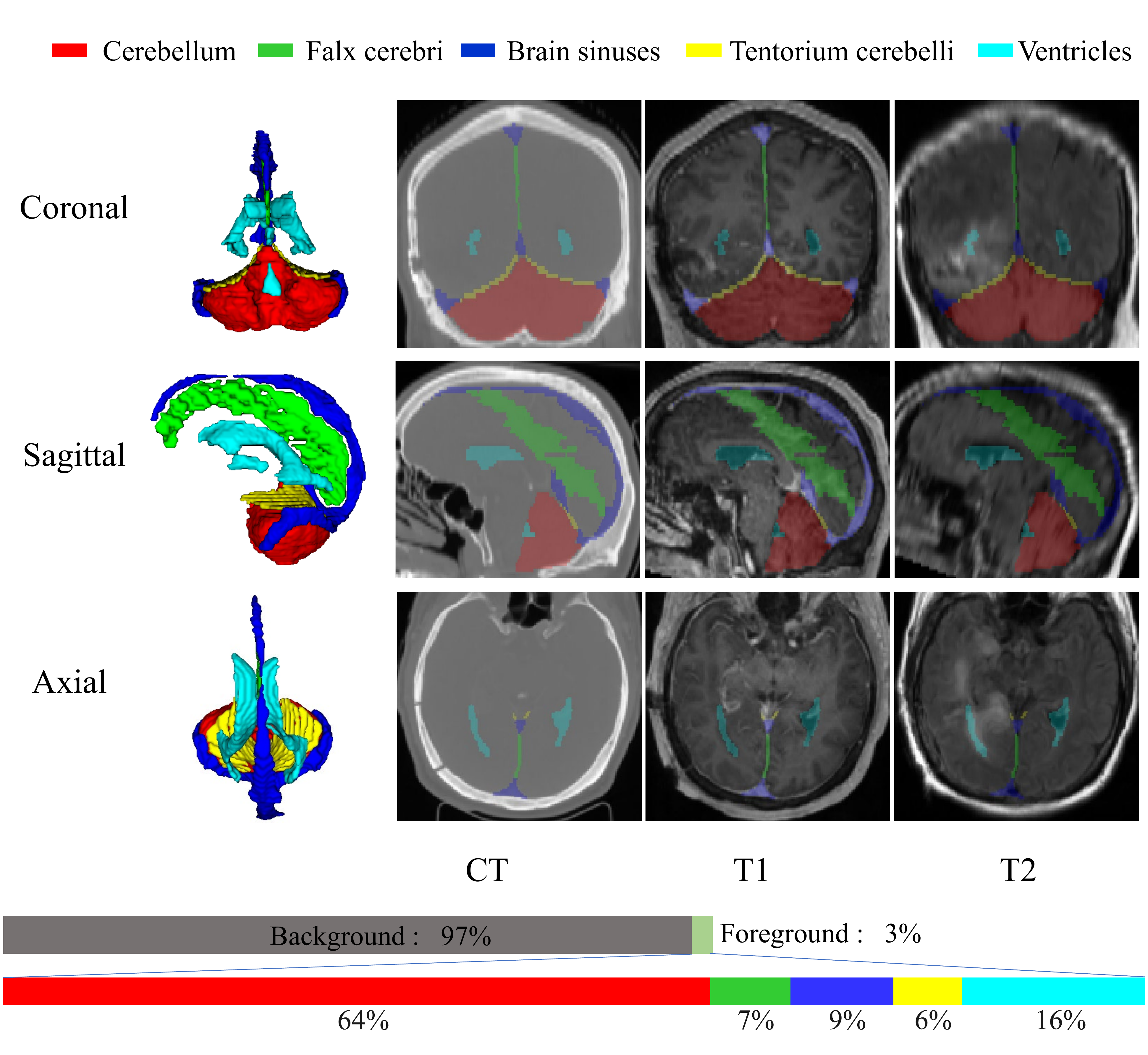}
\caption{Illustration of Anatomical Brain Barriers  to Cancer spread (ABCs). The bars at the bottom show the proportion of the background and different structures. Note that the brain sinuses has a sagittal component and a transverse component. }
\label{fig1:structure description}
\end{figure}

To this end, this work aims to obtain automatic segmentation of ABCs of from medical images, which is important for the accuracy and reasonability of  CTV delineation of  the brain cancer. However, this task is challenging due to several reasons. First, the segmentation involves a set of structures with various shapes and sizes, as shown in Fig.~\ref{fig1:structure description}. While some of the barrier structures are plump in 3D, the shapes of some structures are thin (i.e., slice-like) in 3D space, which makes it hard to use a single 2D or 3D CNNs to deal with all of them well. For example, plump structures like cerebellum and ventricles are more suitable for using 3D CNNs that have an isotropic receptive field in 3D, but typical 3D CNNs are suboptimal for dealing with thin structures like the falx cerebri and brain sinuses that spans only few slices in the sagittal  view. This is due to that the commonly used down- and up-sampling in 3D CNNs would blur the inter-slice information for thin structures. Second, there exists a large imbalance between foreground and background voxels, especially for thin structures. For example, the ratio between tentorium cerebelli and the background is around 1:500. Therefore, taking the entire brain volume as the input of a CNN to segment all the structures simultaneously is unfriendly for such thin structures. Thirdly, the contrast between these structures and surrounding tissues is still low, which can easily lead to under- and over-segmentations. Despite that using multi-modal images can alleviate this problem, it remains challenging to obtain accurate segmentation results of these structures whose shapes and positions can be affected by the presence of tumors.

Recently, various CNNs have been proposed for automatic  medical image segmentation~\cite{ronneberger2015u, milletari2016v, zhou2018unet++,gao2019focusnet, gao2020focusnetv2, lei2021automatic}. One of the most successful CNNs in recent years is the nnU-Net~\cite{isensee2018nnu} that is an extension of 3D U-Net~\cite{cciccek20163d} with an encoder-decoder architecture.  It has been shown good performance in various segmentation tasks including brain tumor segmentation~\cite{isensee2018nnu}. Shusharina \emph{et al.}~\cite{shusharina2020automated} recently applied 3D U-Net~\cite{cciccek20163d} to automatic segmentation of ABCs to assist CTV delineation of brain tumors. However, they used a single CNN to segment all the structures simultaneously without considering the large variance of shapes and positions of structures in this specific task, which limited the segmentation accuracy.

To address the above challenges, we propose a novel framework to segment five ABCs structures shown in Fig.~\ref{fig1:structure description} from multi-modal images. Considering that widely used 3D networks (e.g., 3D U-Net~\cite{cciccek20163d}) can hardly obtain accurate results of multiple structures in one step, we employ such a network to localize each structure roughly, and then use our proposed High and Multi-Resolution network (HMRNet) with Bidirectional Feature Calibration (BFC) blocks to get a fine segmentation of each structure, respectively. 
\par The contribution of this work can be summarized as follows: First, we propose a novel framework for automatic segmentation of anatomical brain barriers to cancer spread from  multi-modality medical images, which has been rarely studied in the literature so far. Second, we propose a novel network structure named as HMRNet that combines high-resolution and multi-resolution branches in parallel to better deal with anatomic structures  with various shapes and scales.  We also introduce two variants of HMRNet (i.e., 3D and 2.5D versions, respectively) to deal with plump and slice-like structures in a unified framework. Thirdly, we introduce a novel attention block that uses bidirectional feature calibration to allow features from different branches to calibrate each other for better feature representation ability. In addition, we propose a novel loss function that can increase the recall and reduce false negatives for segmenting objects with a large imbalance between the foreground and the background. 

\par Our method was originally designed for the task 1 of  MICCAI 2020 ABCs challenge\footnote{https://abcs.mgh.harvard.edu/}. On the public validation dataset, our method won the second place among all the participants\footnote{https://abcs.mgh.harvard.edu/index.php/leader-board}. This work has not been published elsewhere, and in this paper, we provide detailed description of our method and implementation, and validate it with extensive comparison with existing methods and ablation studies.

\section{Related Works}
\subsection{CNNs for Medical Image Segmentation}
Recent years have seen extensive works on CNNs for image segmentation~\cite{litjens2017survey, shen2017deep}. Long \emph{et al.}~\cite{long2015fully} firstly proposed a Fully Convolutional  Network (FCN) that can obtain a full-resolution segmentation of an image in a single forward pass. Ronneberger \emph{et al.}~\cite{ronneberger2015u} proposed U-Net that uses symmetric squeeze and expanding paths to extract features and then recover the spatial resolution of the final prediction map, where skip connections are introduced to fuse shallow and deep features. As an extension of U-Net, U-Net++~\cite{zhou2018unet++} connects the encoder and decoder sub-networks through a series of nested dense skip pathways. V-Net~\cite{milletari2016v} shares a similar structure with U-Net~\cite{ronneberger2015u} and it uses a residual connection in each convolutional block. In~\cite{christ2016automatic}, cascaded FCNs~\cite{brosch2016deep} were used for liver and lesion segmentation.  The nnU-Net~\cite{isensee2018nnu} has achieved state-of-the-art performance in several 3D medical image segmentation tasks. 
Another family of segmentation networks are based on Deeplab~\cite{chen2017deeplab,chen2018encoder}. These networks used dilated convolutions~\cite{yu2015multi} with multiple dilation rates in the same layer to fuse multi-scale information, which will significantly increase the GPU memory consumption, limiting their application to 3D medical image tasks. Recently, Wang \emph{et al.}~\cite{wang2020deep} proposed HRNet that connects high-to-low resolution convolution streams in parallel and repeatedly exchanges the information across resolutions. However, it is also memory consuming and can hardly be used for 3D images.
\par Besides, attention modules have been widely used to strengthen the representation power of CNNs. Hu \emph{et al.}~\cite{hu2018squeeze} proposed the Squeeze-and-Excitation (SE) block to calibrate different channels of a feature map. Rickmann \emph{et al.}~\cite{rickmann2019project} proposed a Project \& Excite (PE) module that squeezes feature maps along different views of a tensor separately to retain more spatial information. To highlight useful regions, attention U-Net~\cite{oktay2018attention} and spatial and channel SE (scSE) networks~\cite{roy2018recalibrating} were proposed for segmentation tasks. The CA-Net~\cite{gu2020net} further combined spatial, channel and scale attentions for segmentation of 2D images. However, these methods like the SE~\cite{hu2018squeeze}  often obtain the calibration coefficients from the input feature map itself, which may be biased and lead to over-fitting. The attention gate proposed in~\cite{oktay2018attention} avoids this problem  by using one high-level feature map to provide the attention signal to calibrate another low-level feature map. However, this is a  monodirectional calibration, and we propose a bidirectional attention in this work that enables feature maps from two different branches to calibrate each other. 
\begin{figure*}
    \centering
    \includegraphics[width=1.9\columnwidth]{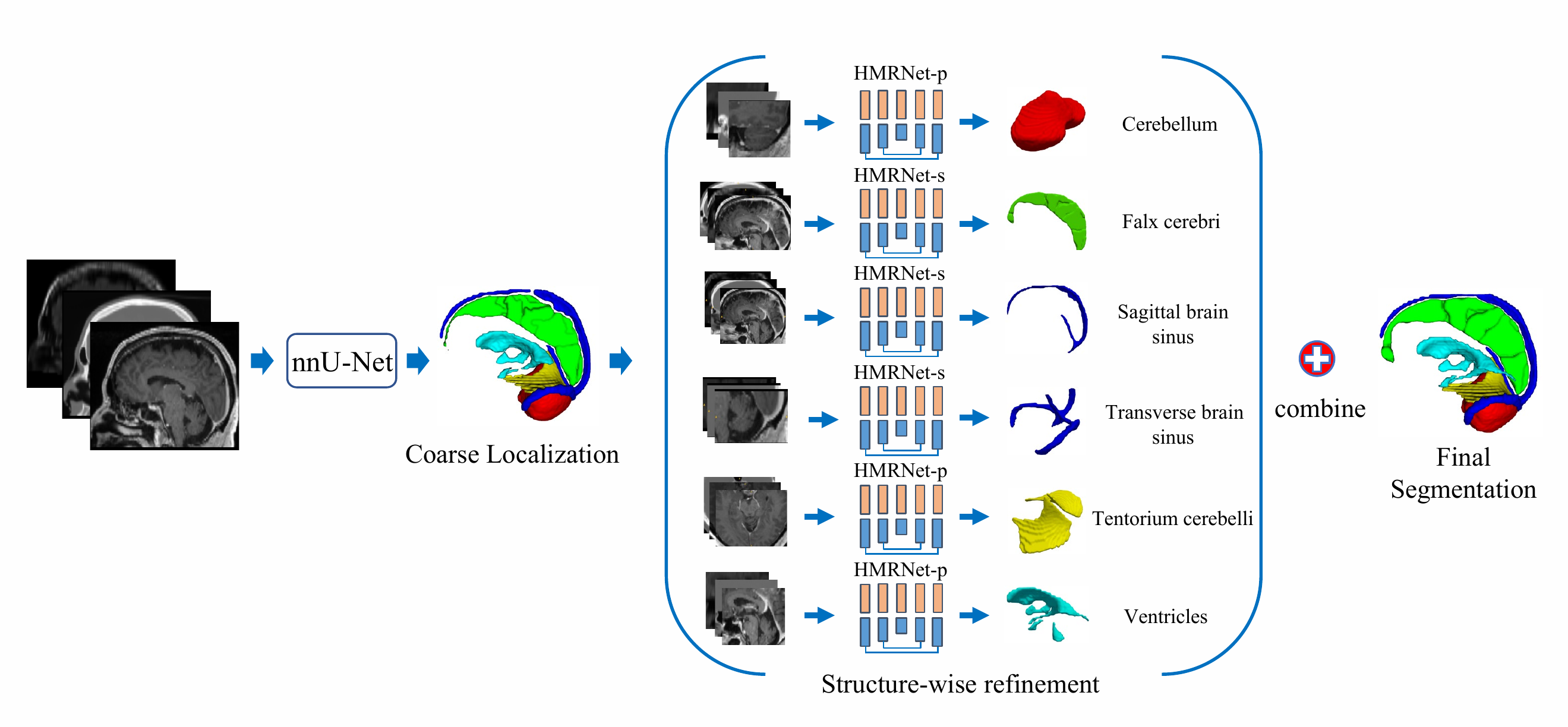}
    \caption{Overall framework of our proposed method for segmentation of ABCs. We first use a coarse localization step to obtain a bounding box of each structure, and then use our proposed HMRNet to obtain a fine segmentation of each structure according to the localization result, respectively. Note that sagittal and transverse brain sinuses are considered respectively, as shown in Fig.~\ref{fig3:organ3_split}. Two variants of HMRNet, i.e., HMRNet-p based on 3D convolutions and HMRNet-s based on 2.5D convolutions, are used to deal with plump and thin structures, respectively.}
    \label{fig2:framework}
\end{figure*}
\subsection{Multi-Organ Segmentation for Radiotherapy}
In early studies, automatic multi-organ segmentation are mainly based on atlases~\cite{glaister2017automatic}. 
For example,  
VaN \emph{et al.}~\cite{van2011automated} segmented hippocampus and cerebellum from MRI based on atlas registration and appearance models. However, these methods require time-consuming image alignment, and the performance is limited when the similarity between the atlas and target images is low. Glaister \emph{et al.}~\cite{glaister2017falx} proposed a multi-atlas approach to segment the falx from T1-weighted MRI and susceptibility-weighted MRI. 

\par Recetnly, deep CNNs have been applied to multi-organ segmentation for radiotherapy planning, such as for Head and Neck (HAN) cancers~\cite{zhu2019anatomynet} and lung cancer \cite{cao2021cascaded}. AnatomyNet~\cite{zhu2019anatomynet} combined 3D U-Net ~\cite{cciccek20163d} and SE block to segment Organs-at-Risks (OARs) for HAN cancers. Considering the imbalance between large and small OARs,  FocusNet \cite{gao2019focusnet} and FocusNet-V2 \cite{gao2020focusnetv2} combined ROI-pooling with  small organ localization and segmentation sub-networks while maintaining accuracy for large organs.  
Lei \emph{et al.}~\cite{lei2021automatic} proposed a 3D SepNet to better deal with thin structures in images with large inter-slice spacing, and it was trained with a hardness-weighted loss for better segmentation. Zou \emph{et al.}~\cite{zou2021domain} applied 3D U-Net~\cite{cciccek20163d} with a label merging strategy for symmetrical organs to segment OARs including the brain stem, chiasm, eyes and optic nerves, etc.
\par Despite the abundant works on OAR segmentation and multi-class segmentation of brain tissues~\cite{chen2018drinet}, methods designed for segmentation of  barrier structure for brain tumors are rare in the literature. To the best of our knowledge, the only two previous works on this task were reported in \cite{shusharina2020automated} and \cite{langhans2021automatic}, where the authors simply employed 3D U-Net ~\cite{cciccek20163d} to segment the barrier structures from multi-modal images, with limited performance  due to the complex shapes and sizes of different imbalanced structures including the falx cerebri, tentorium cerebelli, brain sinuses, cerebellum and ventricles.

\begin{figure}
    \vspace{-0.4cm}
    \centering
    \includegraphics[width=1\columnwidth]{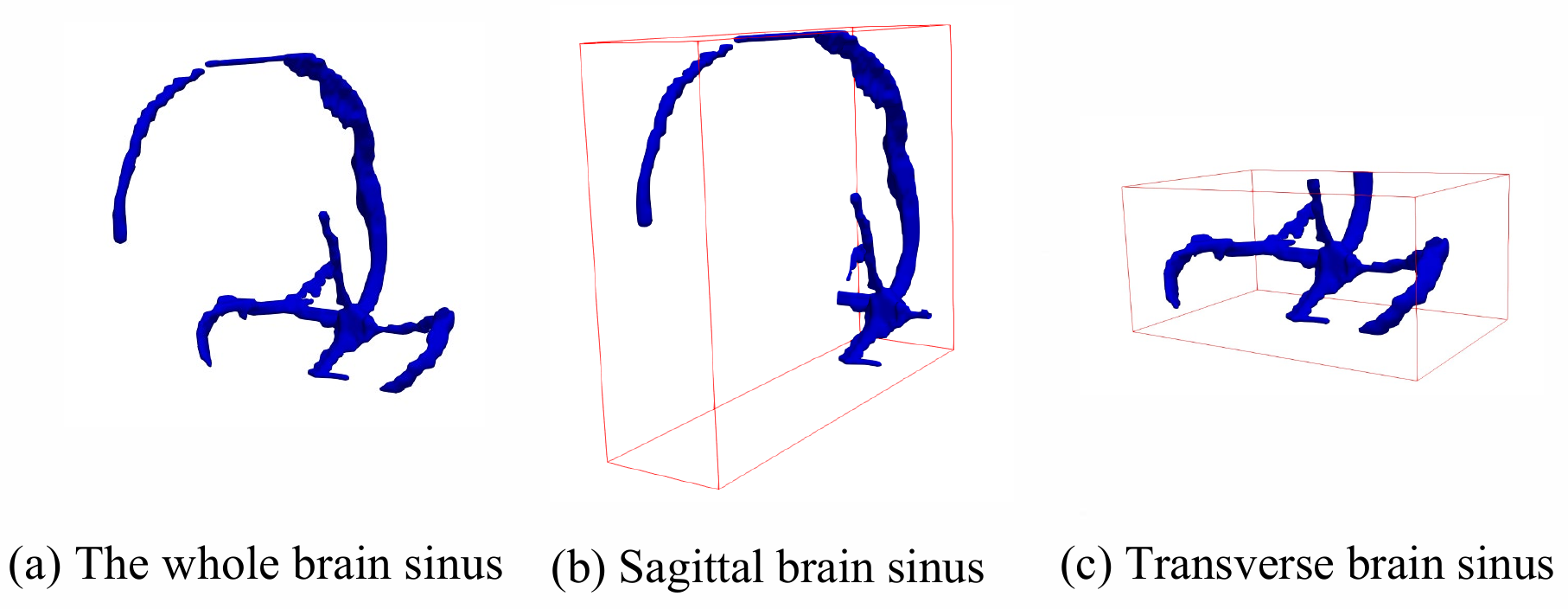}
    \caption{Illustration of the brain sinuses with a “T” shape. Note that the sagittal (b) and transverse (c) components  span few slices in the sagittal view and the axial view respectively. Thus, we use two HMRNet-s to deal with them, respectively.}
    \label{fig3:organ3_split}
\end{figure}
\begin{figure*}
    \centering
    \includegraphics[width=2\columnwidth]{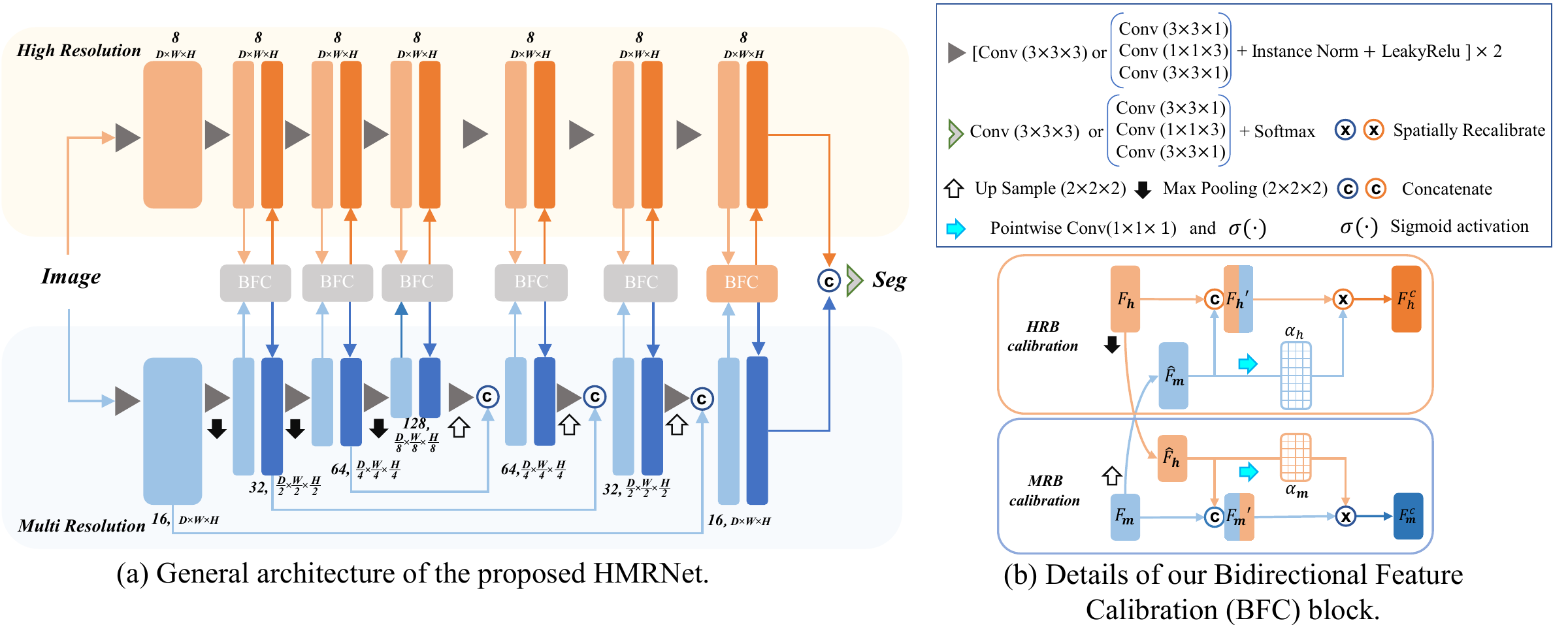}
    \caption{Overview of our proposed HMRNet (a) and BFC (b). The input image is fed into a high-resolution branch to obtain detailed features, and a multi-resolution branch to obtain more abstract features from a larger receptive field. Outputs of the two branches are concatenated to obtain final segmentation prediction. At different depths of the network, features from the two branches are calibrated by each other through the BFC block.  The last orange BFC block does not have  up- and down-sampling operations. Note that HMRNet-p and HMRNet-s are two variants of HMRNet, and they use 3D convolutions and 2.5D convolutions, respectively, where the later is a combination of intra-slice and inter-slice convolutions. }
    \label{fig4:HU-net structure}
\end{figure*}
\section{Methods}

\par Our proposed framework for ABCs segmentation is illustrated in Fig.~\ref{fig2:framework}. As different structures with various shapes and sizes make it hard to segment them simultaneously with a single model, we first use a localization model based on 3D nnU-Net~\cite{isensee2018nnu} to obtain a rough localization of different structures, and then propose a High and Multi-Resolution Network (HMRNet) to segment each structure around its local region respectively, where a Bidirectional Feature Calibration (BFC) block is introduced for better interaction between  features in the two branches. We introduce two variants of HMRNet, i.e., HMRNet-p based on 3D convolution and HMRNet-s based on 2.5D convolution, to deal with plump and thin structures, respectively. A new weighted loss function is also proposed to increase the recall and reduce false negatives for tiny structures. In this section, we introduce our HMRNet with BFC and our proposed loss function, respectively. 

\subsection{High and Multi-Resolution Network (HMRNet)}
The widely used 3D U-Net~\cite{cciccek20163d} for segmenting volumetric medical images employs several down-sampling operations to increase the receptive field and obtain multi-scale features in the encoder, which is helpful for segmentation of relative large structures, but the down-sampling operations lead to poor performance for tiny or thin structures.  
Keeping a high resolution of feature maps is important for obtaining more detailed results of small structures, but simply removing the down-sampling operations makes the network computationally expensive and memory consuming.  

To better deal with the brain barrier structures with both plump and thin structures, we combine the advantages of encoder-decoder-based multi-scale feature learning and high-resolution representation, therefore propose a High and Multi-Resolution Network (HMRNet) that consists of a high-resolution branch and a multi-resolution branch, as shown in Fig.~\ref{fig4:HU-net structure}(a). The high-resolution branch does not use any down- and up-sampling operation, and can keep detailed features for each voxel. The multi-resolution branch is based on a U-shape backbone and able to learn multi-scale features with more abstract representations and contextual information. As the two branches are complementary to each other, we introduce a bidirectional feature calibration block to enable the interaction between the two branches at different depths of the network.  

\par Due to the different shapes of ABCs structures, we design two variants of HMRNet to deal with plump structures (i.e, the cerebellum, the tentorium cerebelli and the ventricles) and slice-like thin structures (i.e., the falx cerebri and the brain sinuses), and the corresponding networks are named as  HMRNet-p and HMRNet-s, respectively. Note that these two variants share the same overall structure, but are implemented with 3D and 2.5D convolutions, respectively.

\textbf{HMRNet-p}. HMRNet-p is designed for plump anatomical structures. As shown in Fig.~\ref{fig4:HU-net structure}(a), the multi-resolution branch of HMRNet-p is a U-shape subnetwork with seven convolutional blocks, which obtains information of different scales through down- and up-sampling. The high-resolution branch also contains seven convolutional blocks while the feature maps' resolution keeps fixed during the whole process. Each convolutional block in both branches contains two $3\times3\times3$ convolutional layers, each followed by an Instance Normalization (IN) layer and a Rectified Linear Units (ReLU) layer. In order to reduce the memory consumption and time cost, we set the number of channels to eight for all blocks in the high-resolution branch. Meanwhile, in the encoder of the multi-resolution branch, the output channel number in the first block is 16 and doubled after each max pooling. The channel number is halved after each block in the decoder. 

At each depth of the network, the corresponding features from the two branches are calibrated by each other through a BFC block, which will be detailed in  Section~\ref{sec:bfc}. Because of the large similarity between feature maps in the first convolution blocks of the two branches, we do not apply BFC in the first convolutional block, as shown in Fig.~\ref{fig4:HU-net structure}(a). The feature maps at the end of the two branches have the same spatial size, and they are concatenated and fed into a $1\times1\times1$ convolutional layer followed by  softmax  to generate the segmentation probabilities. 

\par \textbf{HMRNet-s}. HMRNet-s is designed for slice-like anatomical structures. Due to the special shape of some structures such as the falx cerebri, sagittal and transverse brain sinuses with the slice-like shape that only span few slices,  standard 3D convolutions would blur their boundaries along the through-plane direction and are harmful to the segmentation accuracy. Therefore, we replace a standard 3D convolution layer with a spatially separable convolution that contains two intra-slice convolutions ($3\times3\times1$ kernels)  and one inter-slice convolution ($1\times1\times3$ kernel).
The spatially separable convolutions are also followed by IN and ReLU,
as shown in the legend of Fig.~\ref{fig4:HU-net structure}(a). 

\par Note that the brain sinuses has a spetial “T”-like shape that consists of a thin sagittal component and a thin transverse component, as shown in Fig.~\ref{fig3:organ3_split}. To better deal with this problem, we use two HMRNet-s to segment the sagittal and transverse brain sinuses, respectively, as shown in Fig.~\ref{fig2:framework}.

\subsection{Bidirectional Feature Calibration (BFC)}\label{sec:bfc}
Due to the complementary feature maps from the high-resolution and multi-resolution branches, we propose the BFC block to fuse the corresponding feature maps at the same depth of the two branches and enable a bidirectional calibration between them. In contrast to self-calibration using SE~\cite{hu2018squeeze} that has a risk of over-fitting  and attention gate~\cite{schlemper2019attention,gu2020net} that uses one feature map to monodirectionally calibrate another feature map, our bidirectional calibration makes the two branches aware of each other, which has a potential for more robust learning.
 Let $F_h \in \mathbb{R}^{C_h\times D\times H\times W }$ represent a feature map from the high-resolution branch at a certain depth of the network, where $C_h$ is the channel number and $D\times H\times W $ is the 3D spatial size. We denote the feature map from the multi-resolution branch at the same depth as $F_m \in \mathbb{R}^{C_m\times d\times h\times w }$ with $C_m$ channels and the spatial size of $d\times h \times w$. Note that   $F_m$ has a lower resolution than $F_h$.

As shown in Fig.~\ref{fig4:HU-net structure}(b), the BFC block contains two parts: High-Resolution Branch (HRB) calibration and Multi-Resolution Branch (MRB) calibration. For HRB calibration, we first upsample $F_m$ to the same spatial resolution as $F_h$, and the output is denoted as $\hat{F}_m$ and concatenated with $F_h$:
\begin{equation}\label{eq:emc}
 F'_h = F_h\oplus \hat{F}_m
\end{equation}
where $\oplus$ denotes the concatenation operation, and the output is $ F'_h \in \mathbb{R}^{(C_h + C_m)\times D\times H\times W }$. Then, we squeeze $\hat{F}_m$  along the channel dimension and excite it in the spatial dimension to generate a spatial attention map to calibrate $F'_h$. Specifically, a linear projection matrix $W_m \in \mathbb{R}^{  C_m\times 1\times 1 \times 1} $ squeezes $\hat{F}_m$ into one channel, and then a sigmoid activation $\sigma(\cdot)$ is used to obtain the spatial attention map for HRB, where $W_m$ is implemented by a pointwise convolution. 
\begin{equation}
    \alpha_h = \sigma(W_m \cdot \hat{F}_m)
\end{equation}
where $\alpha_h \in {[0, 1]}^{1\times   H \times W \times D} $ is the spatial attention map for HRB. Each value of $\alpha_h $ indicates the importance of spatial information at the corresponding voxel in $F'_h$. The calibrated feature in HRB is formulated as: 
\begin{equation}
    F^c_h = \alpha_h \cdot F'_h
\end{equation}

Similarly, we use an MRB calibration block to calibrate feature from the multi-resolution branch. We first downsample $F_h$ to the same spatial resolution as $F_m$, and the result is concatenated with $F_m$:
\begin{equation}\label{eq:emc}
 F'_m = F_m\oplus \hat{F}_h
\end{equation}
where $\hat{F}_h$ is the downsampled version of $F_h$, and $F'_m$ is the concatenation output. The attention map is obtained by:
\begin{equation}
    \alpha_m = \sigma(W_h \cdot \hat{F}_h)
\end{equation}
where $\alpha_m \in {[0, 1]}^{1\times   h \times w \times d} $ is the spatial attention map for MRB, and $W_h \in \mathbb{R}^{C_h \times 1 \times 1\times 1}$ is a linear projection matrix implemented by pointwise convolution. The calibrated feature in MRB is formulated as: 
\begin{equation}
    F^c_m = \alpha_m \cdot F'_m
\end{equation}

After using HRB and MRB calibration blocks, the outputs $F^c_h$ and $F^c_m$ in BFC are sent back to the high-resolution branch and multi-resolution branch respectively, which serve as inputs for the corresponding next covolutional blocks in HMRNet, as shown in Fig.~\ref{fig4:HU-net structure}(a). Note that we use BFC at different depths of the HMRNet, leading to six BFC blocks in total.

\subsection{Loss Function}
The large imbalance between the foreground and background voxels for small/thin organs tends to increase false negative predictions. To solve this problem, we design a new loss encouraging a high recall for the fine segmentation. Inspired by the Tversky loss \cite{salehi2017tversky}, we add a hyper-parameter $\theta$ into the Dice loss to control the trade-off between false positives and false negatives. The standard Dice loss and our proposed false negative weighted Dice loss are defined as:
\begin{equation}
    L_{Dice}=1-\frac1{\left|K\right|}\sum_{k=1}^{K}\frac{\sum_i^N2p_i^kg_i^k+\epsilon}{\sum_i^Np_i^k\;+\;\sum_i^Ng_i^k+\epsilon\;}
\end{equation}

\begin{equation}
\begin{aligned}
   L_{fn-Dice}=1 - \frac{\sum_i^N2p_ig_i+\epsilon}{\sum_i^Np_ig_i+\theta\cdot\sum_i^Np_i\overline g_i+\sum_i^N\overline p_ig_i+\epsilon}   
\end{aligned}
\end{equation}
where $N$ and $K$ are the numbers of voxels and classes in the image, respectively.  $\epsilon$ is a small number for numerical stability. $p_i^k$ and $g_i^k$ is the predicted and ground truth probability of pixel $i$ belonging to class $k$. Note that in $L_{fn-Dice}$, we only consider a binary segmentation formulation. $\overline g_i=1-g_i$ and $\overline p_i= 1- p_i$ denote the background probability in the prediction and ground truth, respectively. $p_i\overline g_i$ and $\overline p_ig_i$ denote the False Negative (FN) and the False Positive (FP) predictions, respectively.  $\theta \geq1$ is a hyper-parameter to penalize false negatives in the prediction. In our work, due to the large variance between different structures, an optimal $\theta$ value is selected for each substructure, which will be described in the experiment. The total loss $L_{f-seg}$ for training our HMRNet in the fine segmentation stage is a combination of $L_{fn-Dice}$ and cross entropy loss $L_{CE}$: 
\begin{equation}
\begin{aligned}
    { L}_{f-seg}={ L}_{fn-Dice}+ L_{CE}
\end{aligned}
\end{equation}
where the cross entropy loss is:
\begin{equation}
    { L}_{CE} =-\frac1N{\sum_{i=1}^N}\sum_{k=1}^K g_i^k\log(p_i^k)
\end{equation}

\section{Experiments and results}

\subsection{Experimental Setting}
\subsubsection{Dataset and Evaluation Metrics}\label{sec:data}
We used the dataset of MICCAI 2020 ABCs challenge (task 1) for experiments, which contained images from 45 patients for training, 15 patients for online validation and 15 patients for final online testing. Because the official online testing set was not publicly available, we randomly split the official training data into 36 and 9 for training and testing respectively. The ground truth for the training data was manually annotated by experts. The data of each patient consisted of a CT scan, a contrast-enhanced T1-weighted MRI and a T2-weighted MRI. Images of each patient were registered, and re-sampled to the size of  $164\times194\times142$ with the spacing of $1.2~mm\times1.2~mm\times1.2~mm$ by the organizers. We used z-score~\cite{isensee2018nnu} to normalize the image intensity.  The three modalities for each patient were concatenated into a three-channel volume as the input of CNNs. For quantitative evaluation, we measured the Dice similarity and Average Symmetric Surface Distance (ASSD) between segmentation results and the ground truth in 3D space.

\begin{table*}
\centering{
\caption{{Quantitative evaluations of different modalities. The best result is shown in bold font. Cer: cerebellum, Fal: falx cerebri, Sinu: sagittal \& transverse brain sinuses, Ten: tentorium cerebelli, and Ven: ventricles.}}
\label{table:modality}
\setlength\tabcolsep{2.5pt}
\begin{tabular}{cccccccccccccc} 
\toprule
\multirow{2}{*}{\textbf{Method}}  & \multicolumn{6}{c}{{\textbf{Dice (\%)}}}                                                                            &  & \multicolumn{6}{c}{{\textbf{ASSD} \textbf{(mm)}}}\\ 
\cmidrule{2-7}\cmidrule{9-14}
                        & Cer & Fal & Sinu & Ten & Ven & Average &  & Cer & Fal & Sinu & Ten & Ven & Average  \\ 
\cmidrule{1-7}\cmidrule{9-14}

\multicolumn{1}{l}{{T1}}                  &{ 90.5±2.0}           &{ 75.3±7.9}          &{ 79.8±3.7}          & {74.6±5.8}          &{ 84.1±5.9}          & {80.9±4.8}          &  & {.802±.165}          &{ .561±.154}          &{ .711±.182}          &{ .502±.094}          &{ .725±.130}          &{ .661±.178}             \\
\multicolumn{1}{l}{{T1+T2}}               &{ 93.3±1.3}          &{ 77.5±6.3}          &{ 81.2±2.6}          &{ 77.3±4.9}          &{ 85.3±4.2}          &{ 82.9±3.0}          &  &{ .799±.151}          &{ .541±.142}          &{ .693±.170}          &{ .499±.089}          &{ .708±.113}          &{ .648±.165}               \\
{T1+T2+CT }               &{ \textbf{94.9}$\textbf{±}\textbf{1.0}$}          &{ \textbf{78.6}$\textbf{±}\textbf{5.1}$}          &{\textbf{82.1}$\textbf{±}\textbf{2.4}$}          &{ \textbf{78.0}$\textbf{±}\textbf{4.7}$}          &{ \textbf{87.1}$\textbf{±}\textbf{3.6}$}          &{ \textbf{84.1}$\textbf{±}\textbf{2.5}$}          &  &{ \textbf{.787}$\textbf{±}\textbf{.146}$}          &{ \textbf{.524}$\textbf{±}\textbf{.134}$}          &{ \textbf{.679}$\textbf{±}\textbf{.161}$}          &{ \textbf{.497}$\textbf{±}\textbf{.081}$}          &{\textbf{ .697}$\textbf{±}\textbf{.102}$}          &{ \textbf{.637}$\textbf{±}\textbf{.151}$}          \\
\bottomrule
\end{tabular}}
\end{table*}
\begin{table*}
\centering{
\caption{Quantitative evaluations of different methods for ABCs segmentation. The best result is shown in bold font.  Cer: cerebellum, Fal: falx cerebri, Sinu: sagittal \& transverse brain sinuses, Ten: tentorium cerebelli, and Ven: ventricles.}
\label{table1:coarse_fine_compare}
\setlength\tabcolsep{2.5pt}
\begin{tabular}{ccccccccccccccc} 
\toprule
\multirow{2}{*}{Coarse}            & \multirow{2}{*}{Refine}                                & \multicolumn{6}{c}{\textbf{Dice (\%)}}                                                                                                                                                                                                                                                                   &                      & \multicolumn{6}{c}{\textbf{ASSD} \textbf{(mm)}}                                                                                                                                                                                                                                                            \\ 
\cmidrule{3-8}\cmidrule{10-15}
\multicolumn{1}{c}{}                                   &                                                        & Cer                                             & Fal                                             & Sinu                                           & Ten                                             & Ven                                             & Average                                        &                      & Cer                                            & Fal                                              & Sinu                                   & Ten                                             & Ven                                             & Average                                         \\ 
\cline{1-8}\cline{10-15}
\begin{tabular}[c]{@{}c@{}}nnU-Net\\\cite{isensee2018nnu}\end{tabular}     & ×                                                      & 94.9±1.0                                        & 78.6±5.1                                        & 82.1±2.4                                       & 78.0±4.7                                        & 87.1±3.6                                        & 84.1±2.5                                       &                      & .787±.146                                      & .524±.134                                        & .679±.161                              & .497±.081                                        & .697±.102                                       & .637±.151                                       \\
\begin{tabular}[c]{@{}c@{}}{nnU-Net}\\\cite{isensee2018nnu}{(sub)}\end{tabular} & {×}                                                      & {95.0±1.1}                                        & {78.8±4.9}                                        & {82.1±2.4}                                       & {78.5±4.3}                                        & {88.6±3.6}                                        & {85.1±2.4}                                       &                      & {.643±.137}                                      & {.486±.245}                                        & {.672±.153}                              & {.456±.079}                                        & {.646±.110}                                       & {.602±.103}                                       \\
\begin{tabular}[c]{@{}c@{}}{nnU-Net}\\\cite{isensee2018nnu}\end{tabular}     & \begin{tabular}[c]{@{}c@{}}{nnU-Net}\\\cite{isensee2018nnu}\end{tabular}     & \multicolumn{1}{l}{{95.0±0.8}}                    & \multicolumn{1}{l}{{79.1±4.8}}                    & \multicolumn{1}{l}{{82.4±2.2}}                   & \multicolumn{1}{l}{{79.5±3.7}}                    & \multicolumn{1}{l}{{88.4±3.3}}                    & \multicolumn{1}{l}{{84.8±1.9}}                   & \multicolumn{1}{l}{} & \multicolumn{1}{l}{{.723±.107}}                  & \multicolumn{1}{l}{{.434±.340}}                    & \multicolumn{1}{l}{{.655±.135}}          & \multicolumn{1}{l}{{.433±.085}}                    & \multicolumn{1}{l}{{.638±.113}}                   & \multicolumn{1}{l}{{.576±.132}}                   \\
\begin{tabular}[c]{@{}c@{}}nnU-Net\\\cite{isensee2018nnu}\end{tabular}     & \begin{tabular}[c]{@{}c@{}}nnU-Net\\\cite{isensee2018nnu}(sub)\end{tabular} & 95.3±0.8                                        & 79.3±4.8                                        & 82.6±2.2                                       & 79.9±3.7                                        & 88.6±3.3                                        & 85.1±1.9                                       &                      & .668±.117                                      & .444±.345                                        & .652±.130                     & .385±.077                                        & .614±.101                                       & .573±.127                                       \\
\begin{tabular}[c]{@{}c@{}}nnU-Net\\\cite{isensee2018nnu}\end{tabular} & \multicolumn{1}{l}{HMRNet}                             & \multicolumn{1}{l}{\textbf{95.8}\textbf{±}\textbf{0.7}~} & \multicolumn{1}{l}{\textbf{80.1}\textbf{±}\textbf{4.7~}} & \multicolumn{1}{l}{\textbf{83.5}\textbf{±}\textbf{2.4}} & \multicolumn{1}{l}{\textbf{82.0}\textbf{±}\textbf{3.4~}} & \multicolumn{1}{l}{\textbf{89.5}\textbf{±}\textbf{2.8~}} & \multicolumn{1}{l}{\textbf{86.2}\textbf{±}\textbf{2.0}} & \multicolumn{1}{l}{} & \multicolumn{1}{l}{\textbf{.595}\textbf{±}\textbf{.089}} & \multicolumn{1}{l}{\textbf{.371}\textbf{±}\textbf{.101}~} & \multicolumn{1}{l}{\textbf{.601}\textbf{±}\textbf{.155}} & \multicolumn{1}{l}{\textbf{.353}\textbf{±}\textbf{.070}~} & \multicolumn{1}{l}{\textbf{.576}\textbf{±}\textbf{.052~}} & \multicolumn{1}{l}{\textbf{.499}\textbf{±}\textbf{.081}}  \\
\bottomrule
\end{tabular}}
\end{table*}
\subsubsection{Implementation Details\label{sec:implement}}
The proposed framework was implemented in PyMIC\footnote{https://github.com/HiLab-git/PyMIC} \cite{wang2020noise} using PyTorch on one NVIDIA GTX 2080 Ti GPU with 12 GB memory. We used the SGD optimizer with a weight decay $1\times10^{-5}$ for training. The initial learning rate  was  $\eta_0 = 0.01$, and was decayed based on $\eta=\eta_{0}\cdot\left(1-\frac{e-1}{N_e}\right)^{0.9}$, where $e$ is the current epoch number, and $N_e = 300$ is the total number of epochs. We employed $L_{Dice} + L_{CE}$ for the coarse location task. {The code of our method is available online\footnote{https://github.com/HiLab-git/ABCschallenge2020}}.

To train the fine segmentation models, we cropped the original input volumes for each structure based on the bounding box inferred from the ground truth and expanded by 10 pixels. Random rotation, elastic deformation, gamma correction and random flipping were applied for data augmentation. Deep supervision was used for better performance. {Specifically, we use a pointwise convolution layer at the end of each level of the decoding blocks, and compared it with $G_s$ that is a down-sampled version of the ground truth so that they have the same resolution level. The deep supervision loss is defined as: 
\begin{equation}
\begin{aligned}
L_{deep} = \sum_{s=1}^S \lambda_s L_{f-seg}(P_s, G_s)
\end{aligned}
\end{equation}
where $S = 3$ is the entire number of resolution levels. $\lambda_s$ is the weight of resolution level $s$, and set as 0.3, 0.3 and 0.4 in the experiment, respectively.} During inference, we used the coarse location result to get the bounding box for each structure and also expanded it for cropping, which obtains a subvolume as the input of HMRNet. The optimal values of $\theta$ in $L_{fn-Dice}$ for different structures were determined through five-fold cross validation, and they were: 2.0 for the cerebellum and the transverse brain sinuses,  3.0 for the falx cerebri and the sagittal  brain sinuses, 3.5 for the tentorium cerebelli and 2.5 for the ventricles.

In each iteration for training the second-stage model, we randomly selected the checkpoint of the first-stage model at epoch $e$ ($e$ = 100, 200, 300) to obtain the initial segmentation results for training data, which serves as an augmentation strategy, and the training data for the second-stage model contains both high-quality and low-quality initial segmentation results, which can improve the second-stage model’s tolerance on relatively low-quality initial segmentation.

\begin{figure}
    \centering
    \includegraphics[width=1\columnwidth]{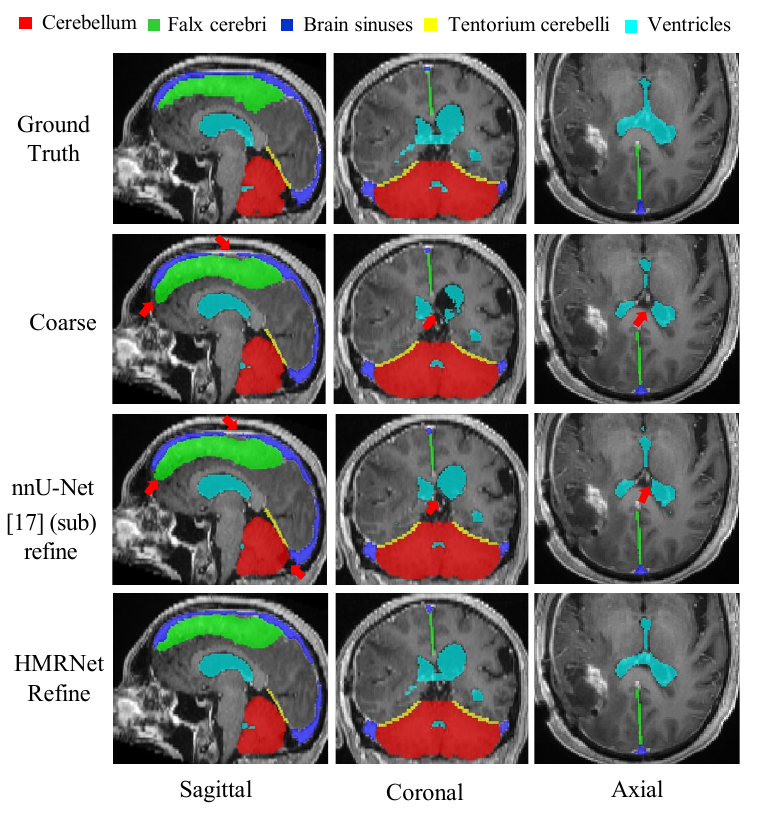}
    \caption{Visual comparison of coarse segmentation and refinement based on nnU-net (sub) and HMRNet, respectively. 
    Red arrows denote the under- and over-segmentation compared with the ground truth. }
    \label{fig5:HU-net structure}
\end{figure}
\begin{table*}
\centering{
\caption{Ablation study of the overall structure of HMRNet when attention mechanisms are not used. HR\_branch and MR\_branch mean only using the HR branch and only  using the MR branch, respectively. HMRNet-p and HMRNet-s represent using HMRNet-p and HMRNet-s for all the structures, respectively. HMRNet denotes using our HMRNet-p for plump structures and HMRNet-s for slice-like structures, respectively.}
\label{table2:effect_of_struc}
\setlength\tabcolsep{3.7pt}
\begin{tabular}{ccccccclcccccc} 
\toprule
\multirow{2}{*}{\textbf{ Network}} & \multicolumn{6}{c}{\textbf{Dice (\%)}}                                                                        &  & \multicolumn{6}{c}{\textbf{ASSD} \textbf{(mm)}}                                                                      \\ 
\cmidrule{2-7}\cmidrule{9-14}
                                   & Cer             & Fal             & Sinu            & Ten             & Ven             & Average         &  & Cer             & Fal             & Sinu            & Ten             & Ven             & Average          \\ 
\cmidrule{1-7}\cmidrule{9-14}
Coarse                             & 94.9±1.0          & 78.6±5.1          & 82.1±2.4          & 78.0±4.7          & 87.1±3.6          & 84.1±2.5          &  & .787±.146          & .524±.134          & .679±.161          & .497±.081          & .697±.102          & .637±.151            \\
HR\_branch                         & 94.9±0.8           & 79.0±4.7           & 82.0±2.5           & 79.7±3.3           & 87.9{±{2.2}}          & 84.7±2.1           &  & .790±.121          & .421±.250           & .714±.213            & .394±.116          & .725±.159            & .609±.123            \\
MR\_branch                         & 95.5±0.8           & 78.9±5.0           & 82.6±2.8           & 79.5±3.5           & 89.1±3.5           & 85.1±1.9           &  & \textbf{.656±.103}           & .436±.338           & .661±.166           & .431±.077           & .617±.157           & .560±.126            \\
HMRNet-p                           & \textbf{95.6±0.6}           & 79.5±4.7           & 82.4±2.3           & 81.4±2.9           & \textbf{89.4±3.0}           & 85.7±1.8           &  & .685±.091           & .360±13.7           & .677±.141           & \textbf{.364±.066}           & \textbf{.578±.061}           & .533±.081            \\
HMRNet-s                           & 95.2±0.7           & 79.8±4.9           & \textbf{82.7±2.4}           & 81.0±3.4           & 89.1±3.3           & 85.6±2.0           &  & .737±.103           & \textbf{.354±.137}           & .654±.156           & .389±{.141}           & .609±.063           & .549±.095            \\
HMRNet                             & 95.6±0.7          & \textbf{79.8±4.7}           & \textbf{82.7}\textbf{±2.4}          & \textbf{81.4}\textbf{±}\textbf{2.8}          & 89.4±3.1           & \textbf{85.8}\textbf{±}\textbf{1.9}          &  & .686±.094           & \textbf{.354}\textbf{±}\textbf{.137}         & \textbf{.649\textbf{±.154}}           & .366±.065          & .581±.061           & \textbf{.527}\textbf{±.086}   \\
\bottomrule
\end{tabular}}
\end{table*}
\subsection{The Role of Different Modalities}
\par {We first investigated the role of different modalities in the segmentation task, and compared using T1, T1 + T2 and T1 + T2 + CT for the segmentation task based on nnU-Net\cite{isensee2018nnu}, respectively. The results are listed in Table~\ref{table:modality}. It can be observed that using T1 only obtained an average Dice of 80.9$\%$ for the five organs, and using the three modalities obtained an average Dice of 84.1$\%$, which outperformed the other two variants. Therefore, it is necessary to combine the thee available modalities for the segmentation task.} 
\subsection{Effectiveness of Our Framework}
\par {We first compared two variants of nnU-Net for segmentation: the original nnU-Net \cite{isensee2018nnu} that segments all the organs simultaneously and nnU-Net (sub) that means segmenting each organ respectively from the entire image using five models of nnU-Net, each was a binary segmentation model. Then using the results of nnU-Net \cite{isensee2018nnu} as coarse segmentation, we compared three different ways for fine segmentation: 1) using nnU-Net~\cite{isensee2018nnu} to refine all the organs simultaneously, 2) using nnU-Net~\cite{isensee2018nnu} (sub)  that means binary nnU-Net \cite{isensee2018nnu} models are used to refine each organ respectively, and 3) using our HMRNet to refine each organ respectively. }Note that in the fine segmentation stage, both the 3D nnU-Net \cite{isensee2018nnu} and HMRNet used the same cropping strategy based on the coarse segmentation, as described in Section~\ref{sec:implement}.  All these methods were trained with $L_{Dice}+L_{CE}$. 

\begin{table*}
\centering{
\caption{Ablation study of BFC used in HMRNet.  $\times$ : without any attention mechanisms, HRB and MRB denote monondirectional feature calibration methods, where HRB uses multi-resolution branch features to calibrate  high-ressolution branch features, and vice versa for MRB. Enco and  Deco represent using our BFC block only in the encoder and the decoder, respectively. {Bottom, Middle and Top represent using BFC block only for the lowest, middle and highest resolution level of the network, respectively. BFC is our proposed bidirectional feature calibration used at all the resolution levels.}  }
\label{tab:attention}
\setlength\tabcolsep{3.5pt}{
\begin{tabular}{ccccccclcccccc} 
\toprule
\multirow{2}{*}{\textbf{ Attention}} & \multicolumn{6}{c}{\textbf{Dice (\%)}}                                                                        &  & \multicolumn{6}{c}{\textbf{ASSD }\textbf{(mm)}}                                                                     \\ 
\cmidrule{2-7}\cmidrule{9-14}
                                     & Cer            & Fal             & Sinu            & Ten             & Ven             & Average         &  & Cer             & Fal             & Sinu            & Ten             & Ven             & Average          \\ 
\cmidrule{1-7}\cmidrule{9-14}
$\times$                                     & 95.6±0.7          & 79.8±4.7           & 82.7±2.4           & 81.4±2.8           & 89.4±3.1           & 85.8±1.9           &  & .686±.094           & .354±.137           & .649±.154           & .366±.065           & .581±.061           & .527±.086            \\
HRB                                  & 95.7±0.7          & 80.0±4.9           & 82.5±2.4           & 81.4±3.1           & 89.4±3.0           & 85.8±1.9           &  & .643±.087           & \textbf{.318}\textbf{±}\textbf{.079}           & .669±.164           & .382±.079           & .617±.136           & .526±{.067}            \\
MRB                                  & 95.4±0.8          & 80.0±4.4           & 83.0±2.6           & 81.0±3.6           & 88.9±3.3           & 85.7±2.0           &  & .712±{.086}           & .421±.306           & .648±.146           & .366±.076           & .603±.080           & .550±.109            \\
Enco                                 & 95.6±0.7          & 79.9±4.8           & 82.7±2.4           & 81.4±{2.4}           & 89.0±2.9           & 85.7±1.9           &  & .685±.091           & .348±.087           & .648±.154           & .355±{.058}           & .611±.123           & .529±.071            \\
Deco                                 & 95.6±{0.6}          & \textbf{80.2}\textbf{±4.9}           & 82.8±2.4           & 81.8±3.5           & 89.3±3.1           & 85.9±2.1           &  & .682± .094           & .406±.305           & .649±.155           & .355±.081           &\textbf{ .550}\textbf{±.077}           & .528±.124            \\
{Bottom}                                 & {95.7±0.9}         & {80.1±4.6}           & {83.1±{2.2}}           & {81.1±3.3}           & {89.4±2.8}           & {85.7±1.9}           &  & {.601±.117}           & {.430±.339}           & {.617±.149}           & {.363±.066}           & {.570±.139}           & {.515±.125}           \\
{Middle}                                 & {95.8±1.0}          & {80.0±4.7}           & {83.1±2.7}           & {80.6±3.7}           & {89.0±3.2}           & {85.7±2.3}           &  & {.621±.130}           & {.431±.342}           & {.66.0±.200}           & {.437±.249}           & {.588±.115}           & {.549±.184}            \\
{Top}                                 & {95.7±0.9}          & {80.1±4.6}           & {83.4±2.3}           & {81.6±3.0}           & {89.4±2.8}           & {86.0±{1.9}}           &  & {.601±.105}           & {.421±.346}           & {.621±{.139}}           &{.373±.064}            & {.595±.060}           & {.514±.143}            \\
BFC                                  & \textbf{95.8\textbf{±0.7}}           & 80.1±{4.4}           & \textbf{83.5\textbf{±2.4}}          & \textbf{82.0\textbf{±3.4}}          & \textbf{89.5}\textbf{±}\textbf{2.8}           & \textbf{86.2\textbf{±2.0}}           &  & \textbf{.595\textbf{±.089}}            & .371{±.101}          & \textbf{.601\textbf{±.155}}          &\textbf{.353\textbf{±.070}}           & .576±{.052}          & \textbf{.499\textbf{±.081}} \\
\bottomrule
\end{tabular}}}
\end{table*}
Quantitative evaluation results in Table~\ref{table1:coarse_fine_compare} show that using another nnU-Net for refinement can slightly improve the average Dice from 84.1\% to 84.8\%, but using nnU-Net (sub) and HMRNet for refinement can improve the avaerge Dice to 85.1\% and 86.2\%, respectively. Using HRMNet to refine the coarse segmentation of nnU-Net outperformed the other variants. Compared with the coarse segmentation of nnU-Net, our proposed HMRNet reduced the average ASSD from 0.637 mm to 0.499 mm, which is a 22\% improvement.
A visual comparison of coarse segmentation, nnU-net \cite{isensee2018nnu}(sub) refine and
HMRNet refine is shown in Fig.~\ref{fig5:HU-net structure}. 
From the sagittal view, it could be observed that the prediction of HMRNet has better connectivity than the other two methods in thin structures like sagittal \& transverse brain sinuses and higher completeness in ventricles, which is highlighted by red arrows.

\subsection{Ablation Study of HMRNet}
First, to validate the overall structure of HMRNet, we compared it with four variants: 1) ``HR\_branch" that means only using the high-resolution branch in our HMRNet, 2) ``MR\_branch" that means only applying the multi-resolution branch in HMRNet, 
3) ``HMRNet-p" that means using HMRNet-p for all the structures, and 4) ``HMRNet-s" that means using HMRNet-s for all the structures. Note that our ``HMRNet" method uses HMRNet-p for plump structures (cerebellum, tentorium cerebelli and ventricles) and HMRNet-s for slice-like structures (falx cerebri,  sagittal and transverse brain sinuses), respectively. All these methods were trained with  $L_{Dice}+L_{CE}$ loss for each structure, and we disabled BFC that will be analyzed in the following. As shown in Table~\ref{table2:effect_of_struc}, ``HR\_branch" and ``MR\_branch" achieved lower performance than the other variants. ``HMRNet-p" outperformed  ``HMRNet-s" for the cerebellum, tentorium cerebelli and venricles that are plump structures, and performed worse than ``HMRNet-s" for the falx cerebri and the brain sinuses that are slice-like structures. ``HMRNet" outperformed all the compared variants, which demonstrates the effectiveness of combining the two branches and using different types of convolutoins for the plump and slice-like structures. 
\begin{table*}
\centering{
\setlength\tabcolsep{3.4pt}
\caption{Effect of $\theta$ in our false negative-weighted loss function. Note that when $\theta=1.0$, our $L_{fn-Dice}$ is equal to $L_{Dice}$. }
\label{table4:com_loss}
\begin{tabular}{ccccccccccccccc} 
\toprule
\multirow{2}{*}{$\theta$} & \multirow{2}{*}{} & \multicolumn{6}{c}{\textbf{Dice (\%)}}                                                                        &  & \multicolumn{6}{c}{\textbf{ASSD  (mm)}}                                                                      \\ 
\cmidrule{3-8}\cmidrule{10-15}
                             &                   & Cer            & Fal             & Sinu            & Ten             & Ven             & Average         &  & Cer             & Fal             & Sinu            & Ten             & Ven             & Average          \\ 
\cmidrule{1-8}\cmidrule{10-15}
1.0                            &                   & 95.8±0.7           & 80.1±4.7            & 83.5±2.4            & 82.0±3.4            & 89.5±2.8            & 86.2±2.0            &  & .595±{.089}            & .371±.101           & .601±.155           &.353±.070           & .576±{.052}           & .499±.081            \\
1.5                          &                   & 95.9±0.8           & 80.0±4.8            & 84.0±{2.3}            & 82.1±3.1            & 89.4±2.9            & 86.3±1.9            &  & .618±.106            & .372±.161            & .588±{.108}            & .355±{.057}            & .575±.072            & .502±.081             \\
2.0                            &                   & 96.1±1.0           & 80.1±4.8            & 84.1±2.5            & 82.5±3.2            & 89.6±3.1            & 86.5±1.9            &  & .590±.134            & .341±.098            & .584±.115            & .352±.070            & .565±.061            & .486±.075             \\
Adaptive                     &                   & \textbf{96.1\textbf{±1.0}} & \textbf{80.4}\textbf{±}\textbf{4.4} & \textbf{84.3\textbf{±2.6}} & \textbf{82.8}\textbf{±}\textbf{2.5} & \textbf{89.9}\textbf{±}\textbf{2.7} & \textbf{86.7}\textbf{±}\textbf{1.7}  &  &  \textbf{.589\textbf{±.135}} & \textbf{.311}\textbf{±}\textbf{.064} & \textbf{.577\textbf{±.120}} & \textbf{.347\textbf{±.065}} & \textbf{.557\textbf{±.057}} & \textbf{.476}\textbf{±}\textbf{.074}  \\
\bottomrule
\end{tabular}}
\end{table*}
\begin{table*}
\centering{
\caption{Quantitative evaluation of different methods for segmentation of ABCs. $^*$ denotes significant improvement from the existing methods ($p$-value $<$ 0.05).  
}
\label{table5:all_com}
\setlength\tabcolsep{1.6pt}
\footnotesize
\begin{tabular}{cccccccclcccccc}
\toprule
\multirow{2}{*}{\textbf{ Method}} &  & \multicolumn{6}{c}{\textbf{Dice (\%)}}                                                                        &  & \multicolumn{6}{c}{\textbf{ASSD  (mm)}}                                                                          \\ 
\cmidrule{3-8}\cmidrule{10-15}
                                  &  & Cer             & Fal             &  Sinu             & Ten             & Ven             & Average         &  & Cer             & Fal             &  Sinu             & Ten             & Ven             & Average          \\ 
\cmidrule{1-8}\cmidrule{10-15}
3D U-Net \cite{cciccek20163d}                              &  & 88.3±4.2           & 72.3±6.5            & 76.0±3.8           & 70.1±7.8            & 85.6±4.3           & 78.5±3.8           &  & 1.634±.473            & .516±.252           & .878±.217           & .611±.219           & 2.026±1.075           & 1.133±.191            \\
3D Res U-Net \cite{diakogiannis2020resunet}                     &  & 94.3±0.9           & 75.6±4.9           & 80.0±3.1           & 75.0±6.3           & 84.7±3.4           & 81.9±2.6           &  & .877±.135           & .538±.321           & .763±.192           & .509±.155           & .815±.129           & .700±.133            \\
U-Net cSE \cite{hu2018squeeze}                           &  & 93.3±3.7           & 79.0±5.2           & 81.7±3.1           & 78.0±6.1           & 81.6±6.2           & 82.7±3.9           &  & 1.055±.555           & .467±.171           & .690±.204           & .437±.172           & 1.810±.384           & .892±.239            \\
U-Net sSE \cite{roy2018concurrent}                          &  & 94.6±1.6            & 77.4±6.0            & 80.4±4.0            & 76.7±9.1            & 76.8±17.2            & 81.2±6.8            &  & .875±.301            & .438±.225            & .698±.186            & .476±.238            & 1.286±.389            & .755±.268             \\
U-Net scSE \cite{roy2018recalibrating}                         &  & 94.1±1.9           & 76.6±8.5           & 81.4±3.0           & 76.7±8.1           & 85.4±3.3           & 82.8±4.1           &  & .919±.310           & .419±.191           & .733±.198           & .476±.223           & 1.458±.989           & .801±.318            \\
nnU-Net \cite{isensee2018nnu}                            &  & 94.9±0.6           & 78.6±4.9           & 82.1±2.4           & 78.0±5.1           & 87.1±2.9           & 84.1±1.8           &  &.787±.112           & .524±.079           & .679±.136           & .497±.106           & .697±.095           & .637±.074            \\
FocusNet \cite{gao2019focusnet}                          &  & 94.6±0.9            & 78.5±6.6            & 81.2±2.6            & 76.5±5.4            & 87.5±3.4            & 83.7±2.4            &  & .838±.124            & .354±.137            & .695±.138            & .457±.117            & .770±.245            & .623±.108             \\
3D SepNet \cite{lei2021automatic}                          &  & 95.3±{0.6}           & 79.5±4.5           & 81.8±2.9           & 78.5±5.3           & 87.5±3.3           & 84.5±2.3           &  & .710±.098           & .316±.079           & .708±.200           & .419±.117           & .693±.110           & .569±.101            \\
Ours ($\theta = 1.0$)                           &                   & 95.8±0.7           & 80.1±4.7            & 83.5±{2.4}            & 82.0±3.4            & 89.5±2.8            & 86.2±2.0            &  & .595±{.089}            & .371±.101           & .601±.155           &.353±.070           & .576±{.052}           & .499±.081            \\
Ours                            &  & \textbf{96.1}$\textbf{±0.7}^*$ & \textbf{80.4}$\textbf{±}\textbf{4.4}^*$ & \textbf{84.3$\textbf{±2.6}^*$} & \textbf{82.8}$\textbf{±}\textbf{2.5}^*$ & \textbf{89.9}$\textbf{±}\textbf{2.7}^*$ & \textbf{86.7}$\textbf{±}\textbf{1.7}^*$ &  & \textbf{.589$\textbf{±.135}^*$} & \textbf{.311}$\textbf{±}\textbf{.064}$ & \textbf{.577}$\textbf{±}\textbf{.120}^*$ & \textbf{.347}$\textbf{±}\textbf{.065}$ & \textbf{.557$\textbf{±.057}^*$} & \textbf{.476}$\textbf{±}\textbf{.074}^*$  \\
\bottomrule
\end{tabular}}
\end{table*}
\begin{figure*}[!]
    \centering
    \includegraphics[width=1\linewidth]{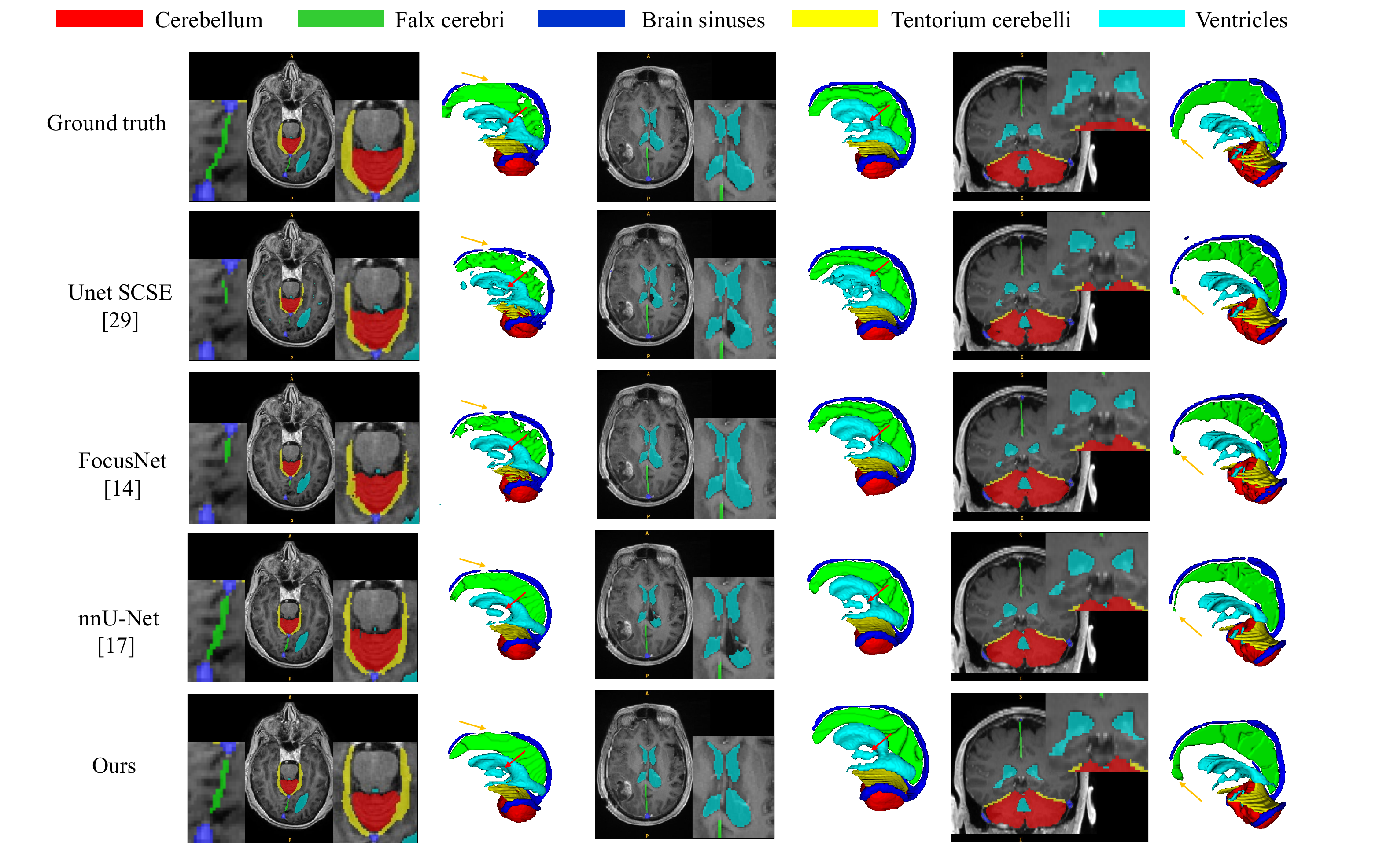}
    \caption{Visual comparison of segmentation results obtained by different methods, which contains three cases with the 2D visualizations in odd columns and 3D visualizations in even columns. Yellow and red arrows highlight the difference of local segmentation results.}
    \label{fig6:all_experients_comparision}
\end{figure*}
\par Then, to validate the effectiveness of our attention mechanism using BFC, we compared it with eight variants: 1) “$\times$” that means not using any attention mechanism, 2) ``HRB" that means only using the HRB calibration, 3) ``MRB" that means only using the MRB calibration, 4) ``Enco" that represents using our BFC only in the encoder blocks, 5) ``Deco” that represents using our BFC only in the decoder blocks, { 6) ``Bottom” that means using the BFC in the smallest resolution level of the multi-resolution branch (channel number 128), 7) ``Middle” that represents using BFC only in the middle resolution level of the multi-resolution branch (channel number 64), and 8) ``Top” that means that using BFC only in the highest resolution level of the multi-resolution branch (channel number 16).} Note that ``HRB" and ``MRB" are monodirectional feature calibration, and our proposed BFC is bidirectional feature calibration and used at all the convolutoinal blocks except the first block. All these methods were trained with the normal $L_{Dice}+L_{CE}$ loss for each substructure. Quantitative evaluation results of these methods are listed in Table~\ref{tab:attention}, which shows that compared with not using attention, our BFC improved the average Dice score from 85.8$\%$ to 86.2$\%$, and reduced average ASSD from 0.527 mm to 0.499 mm, respectively. The proposed BFC also outperformed the other counterparts, which demonstrates the superiority of our bidirectional feature calibration compared with monodirectional calibration methods. 

\subsection{Effectiveness of Our Weighted Dice Loss }
\par To investigate the performance of our false negative-weighted Dice loss $L_{fn-Dice}$ combined with $L_{CE}$, we compared different strategies of setting the value of $\theta$: 1) $\theta = 1.0$ that is equal to using typical $L_{Dice} + L_{CE}$, 2) $\theta = 1.5$ for all structures, 3) $\theta = 2.0$ for all structures, and 4) “adaptive” that refers to using adaptive $\theta$ values for different structures as described in Section~\ref{sec:implement}. Results in Table~\ref{table4:com_loss} show that $\theta = 1.5$ and $\theta = 2.0$ outperformed $\theta = 1.0$, which demonstrates that assigning higher weights to false negatives could improve the performance. Table~\ref{table4:com_loss} also shows that making $\theta$ adaptive to different structures performed better than using the same $\theta$ value for all the structures.

\subsection{Comparison with State-of-the-art Methods}
\par We compared our proposed framework with two recent methods designed for multiple OAR segmentation from head and neck images: FocusNet~\cite{gao2019focusnet} that is an end-to-end two-stage CNN adopting a segmentation-by-detection strategy, and 3D~SepNet~\cite{lei2021automatic} that applies anisotropic convolutions to segment multiple organs in images with large inter-slice spacing. They were also compared with six state-of-the-art networks, i.e., 3D U-Net~\cite{cciccek20163d}, nnU-Net~\cite{isensee2018nnu}, 3D Res~U-Net~\cite{diakogiannis2020resunet}, U-Net~cSE~\cite {hu2018squeeze} that combines 3D U-Net with spatial squeeze and channel excitation block, U-Net~sSE~\cite{roy2018concurrent} that combines 3D U-Net with channel squeeze and spatial excitation block, and U-Net~scSE~\cite{roy2018recalibrating} that combines 3D U-Net with concurrent spatial and channel squeeze and excitation block. All the alternative methods were trained with  $L_{Dice} +  L_{CE}$. Our method using $L_{Dice} + L_{CE}$ in the first stage and $L_{f-seg}$ in the second stage was also compared with using $L_{Dice} + L_{CE}$ for both two stages, which is referred to as “Ours ($\theta = 1.0$)”.

\par Table~\ref{table5:all_com} shows quantitative comparison of these methods on the same testing set. It can be observed that 3D U-Net~\cite{cciccek20163d} got the worst performance in average, and the results of U-Net cSE, U-Net sSE and U-Net scSE demonstrate that using attention networks helped to improve the segmentation accuracy.  FocusNet~\cite{gao2019focusnet} and 3D SepNet~\cite{lei2021automatic} performed better than the other existing methods, but they are inferior to our method. It should be noted that “Ours ($\theta = 1.0$)” already outperformed existing methods when trained with the same loss function, and using our $L_{fn-Dice}$ further improved the performance, which achieved an average Dice and ASSD of 86.7$\%$ and 0.476 mm, respectively. We also did a t-test between our proposed method and the state-of-the-art methods. For the average Dice, the corresponding p-values were 0.0007(3D U-Net~\cite{cciccek20163d}), 0.0003(3D Res U-Net \cite{diakogiannis2020resunet}) , 0.0106 ( U-Net~cSE~\cite {hu2018squeeze}), 0.0285 (U-Net~sSE~\cite{roy2018concurrent}), 0.0144 (U-Net~scSE~\cite{roy2018recalibrating}), 0.0092 (nnU-Net~\cite{isensee2018nnu}), 0.0003 (FocusNet~\cite{gao2019focusnet}) and 0.0021 (3D SepNet~\cite{lei2021automatic}), respectively. All the p-values were less than 0.05, indicating that our proposed method has a significant improvement, as shown in Table~\ref{table5:all_com}. Fig.~\ref{fig6:all_experients_comparision} shows a visual comparison of these methods. It can be observed that our method can better deal with thin structures like the falx cerebri and the brain sinuses than the others, as highlighted by yellow and red arrows.

\section{Discussion and Conclusion}
 {In this paper, we focus on segmentation of five brain organs for radiotherapy planning. The method was specifically designed for these organs due to their special shapes. It would be of interest to investigate the performance of our method when applied to other segmentation tasks and new data in the future. In addition, a general problem for multi-modal segmentation is the requirement of a registration between the different modalities. In this work, the three modalities had been registered by the ABCs challenge organizers, so we did not have an extra registration step for preprocessing.  When our method is applied to a new segmentation task, a spatial alignment for data preprocessing is still necessary, the quality of which may have an impact on the segmentation results. However, for brain structures, rigid registration is usually enough to achieve good results for different modalities of the same patient \cite{de2012multi}.}
 \par  {There are still some limitations of this work. First, our high performance is obtained at the cost of inference speed. Our two-stage method is less efficient than one-stage segmentation methods. However, in the refine stage, each refine model only deals with a subregion of the image, which is fast. Currently, our refine step runs for each organ sequentially, and the entire inference time is 285s, compared with 152s for nn-UNet \cite{isensee2018nnu} and 106s for 3D U-Net~\cite{cciccek20163d}. If the refinement step for each organ runs in parallel, the inference time can be 201s, which is only 48s slower than nnU-Net\cite{isensee2018nnu}. The number of parameters of HMRNet is 3.4 million, compared with 2.7 million of U-Net, and the  Multiply–Accumulate Operations (MACs) of HMRNet and U-Net are 234.15 GMac and 116.45 GMac respectively. However, due to the large variance of shapes and sizes of different organs, the two-stage method is more suitable for small organs. Second, the network structure is designed manually which maybe sub-optimal, it is of interest to use Network Architecture Search (NAS) \cite{guo2020organ}, \cite{liu2020scam} to find better networks for the segmentation task in the future.
\par In summary, we propose a novel framework for the segmentation of anatomical brain barriers to cancer spread from multi-modal brain images. To deal with multiple structures with different sizes and shapes, we propose HMRNet that combines a high-resolution branch and a multi-resolution branch to leverage different contextual information. Two variants of HMRNet, HMRNet-p and HMRNet-s, are proposed for dealing with plump-like and slice-like structures, respectively. We also introduce bidirectional feature calibration in  HMRNet to enable interaction and mutual calibration between features from the two branches. Our framework is built on the top of the state-of-the-art nnU-Net~\cite{isensee2018nnu} to obtain a coarse localization result, and further uses our HMRNet trained with a false negative weighted loss to achieve leading performance on the MICCAI 2020 ABCs challenge dataset. In the future, it is of interest to apply our HMRNet to other medical image segmentation tasks.  


\bibliographystyle{IEEEtran}
\bibliography{reference/refer.bib}
\end{document}